\documentclass[fleqn,usenatbib]{mnras}

\usepackage{newtxtext,newtxmath}

\usepackage[T1]{fontenc}

\DeclareRobustCommand{\VAN}[3]{#2}
\let\VANthebibliography\thebibliography
\def\thebibliography{\DeclareRobustCommand{\VAN}[3]{##3}\VANthebibliography}

\usepackage{graphicx}	
\usepackage{adjustbox}  
\usepackage{amsmath}	
\usepackage{wasysym}    
\usepackage{bm}         
\usepackage{xspace}
\makeatletter 
  \patchcmd{\NAT@citex}
    {\@citea\NAT@hyper@{%
      \NAT@nmfmt{\NAT@nm}%
      \hyper@natlinkbreak{\NAT@aysep\NAT@spacechar}{\@citeb\@extra@b@citeb}%
      \NAT@date}}
    {\@citea\NAT@nmfmt{\NAT@nm}%
    \NAT@aysep\NAT@spacechar\NAT@hyper@{\NAT@date}}{}{}

  \patchcmd{\NAT@citex}
    {\@citea\NAT@hyper@{%
      \NAT@nmfmt{\NAT@nm}%
      \hyper@natlinkbreak{\NAT@spacechar\NAT@@open\if*#1*\else#1\NAT@spacechar\fi}%
        {\@citeb\@extra@b@citeb}%
      \NAT@date}}
    {\@citea\NAT@nmfmt{\NAT@nm}%
    \NAT@spacechar\NAT@@open\if*#1*\else#1\NAT@spacechar\fi\NAT@hyper@{\NAT@date}}
    {}{}
\makeatother

\newcommand\Msun{\text{M}_{\astrosun}} 
\newcommand\HI{\ion{H}{I}\xspace} 
\newcommand\HII{\ion{H}{II}\xspace} 
\newcommand\thesan{\mbox{\textsc{thesan}}\xspace}
\newcommand\thesanone{\mbox{\textsc{thesan-1}}\xspace}
\newcommand\thesantwo{\mbox{\textsc{thesan-2}}\xspace}
\newcommand\thesanwc{\mbox{\textsc{thesan-wc-2}}\xspace}

\newcommand\areport{\mbox{\textsc{arepo-rt}}\xspace}

\newcommand\zreion{$z_{\text{reion}}$\xspace}

\newcommand\orcid[1]{\href{http://orcid.org/#1}{\adjustbox{trim={-.15\width} {0\height} {-.15\width} {0\height},clip}{\includegraphics[height=12pt]{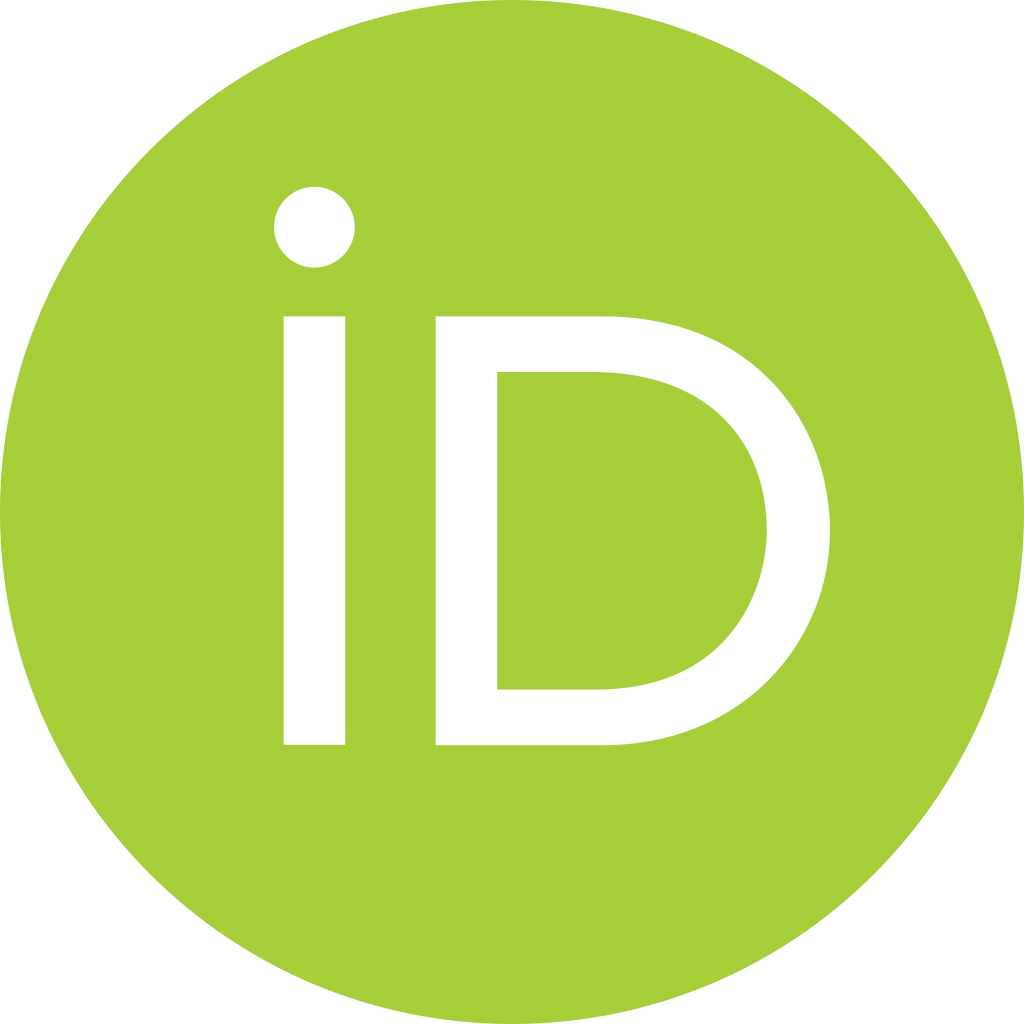}}}}

\title[Ionized bubble sizes during the EoR]{The \textsc{thesan} project: connecting ionized bubble sizes to their local environments during the Epoch of Reionization}

\author[M.~Neyer et al.]{%
Meredith~Neyer\orcid{0000-0002-9205-9717},$^{1}$\thanks{E-mail: \href{mailto:mneyer@mit.edu}{mneyer@mit.edu}}
Aaron~Smith\orcid{0000-0002-2838-9033},$^{2}$
Rahul~Kannan\orcid{0000-0001-6092-2187},$^{3}$
Mark~Vogelsberger\orcid{0000-0001-8593-7692},$^{1,4}$
\newauthor
Enrico~Garaldi\orcid{0000-0002-6021-7020},$^{5}$
Daniela~Gal\'{a}rraga-Espinosa\orcid{0000-0002-8808-803X},$^{5}$
Josh~Borrow\orcid{0000-0002-1327-1921},$^{1,6}$
\newauthor
Lars~Hernquist,$^{7}$
R\"{u}diger~Pakmor\orcid{0000-0003-3308-2420}$^{5}$
and
Volker~Springel\orcid{0000-0001-5976-4599}$^{5}$
\\%
\\%
$^{1}$Department of Physics $\&$ Kavli Institute for Astrophysics and Space Research, Massachusetts Institute of Technology, Cambridge, MA 02139, USA\\%
$^{2}$Department of Physics, The University of Texas at Dallas, Richardson, Texas 75080, USA\\%
$^{3}$Department of Physics and Astronomy, York University, 4700 Keele Street, Toronto, ON M3J 1P3, Canada\\%
$^{4}$The NSF AI Institute for Artificial Intelligence and Fundamental Interactions, Massachusetts Institute of Technology, Cambridge, MA 02139, USA\\%
$^{5}$Max-Planck Institute for Astrophysics, Karl-Schwarzschild-Str.~1, D-85741 Garching, Germany\\%
$^{6}$Department of Physics and Astronomy, University of Pennsylvania, 209 South 33rd Street, Philadelphia, PA, USA 19104\\%
$^{7}$Center for Astrophysics $\vert$ Harvard $\&$ Smithsonian, 60 Garden Street, Cambridge, MA 02138, USA%
}

\date{Accepted XXX. Received YYY; in original form ZZZ}

\pubyear{2024}

\begin{document}
\label{firstpage}
\pagerange{\pageref{firstpage}--\pageref{lastpage}}
\maketitle

\begin{abstract}
An important characteristic of cosmic hydrogen reionization is the growth of ionized gas bubbles surrounding early luminous objects. Ionized bubble sizes are beginning to be probed using Lyman-$\alpha$ emission from high-redshift galaxies, and will also be probed by upcoming 21-cm maps. We present results from a study of bubble sizes using the state-of-the-art \thesan radiation-hydrodynamics simulation suite, which self-consistently models radiation transport and realistic galaxy formation. We employ the mean-free path method, and track the evolution of the effective ionized bubble size at each point ($R_{\rm eff}$) throughout the Epoch of Reionization. We show there is a slow growth period for regions ionized early, but a rapid ``flash ionization'' process for regions ionized later as they immediately enter a large, pre-existing bubble. We also find that bright sources are preferentially in larger bubbles, and find consistency with recent observational constraints at $z \gtrsim 9$, but tension with idealized Lyman-$\alpha$ damping-wing models at $z \approx 7$. We find that high overdensity regions have larger characteristic bubble sizes, but the correlation decreases as reionization progresses, likely due to runaway formation of large percolated bubbles. Finally, we compare the redshift at which a region transitions from neutral to ionized (\zreion) with the time it takes to reach a given bubble size and conclude that \zreion is a reasonable local probe of small-scale bubble size statistics ($R_\text{eff} \lesssim 1\,\rm{cMpc}$). However, for larger bubbles, the correspondence between \zreion and size statistics weakens due to the time delay between the onset of reionization and the expansion of large bubbles, particularly at high redshifts.
\end{abstract}




\section{Introduction}

Hundreds of millions of years after the Big Bang, the first stars and galaxies began to form and emit radiation that ionized the surrounding hydrogen gas to mark the end of the cosmic dark ages and the beginning of the Epoch of Reionization \citep[EoR;][]{Shapiro1987, BarkanaLoeb2001, Furlanetto2006b, Wise2019}. During the EoR, which took place at redshifts of $z \approx 5 - 20$, the gas in the Universe underwent a phase transition during which ionized bubbles expanded around radiation sources, eventually yielding a nearly fully ionized Universe as observed today. Central to this process is the percolation of ionized bubbles as they merged into one another causing rapid increases in bubble sizes \citep{Furlanetto2016}. Observing galaxies and the intergalactic medium (IGM) during the EoR poses challenges due to its high redshift. However, recent and forthcoming \textit{James Webb Space Telescope (JWST)} surveys and 21\,cm telescopes promise insights into high-redshift galaxies and ionized bubble properties, respectively \citep{Robertson2022}. A complete picture of the EoR requires understanding both the IGM and galaxy sources, motivating studies to decode ionized bubble dynamics and morphologies as well as their connection to the local environment \citep{Gnedin2022}. 

The EoR can be explored using a variety of observational techniques including analyzing spectra of Lyman $\alpha$ emitting galaxies at high redshift. The recently-launched \textit{JWST} is revolutionizing the field of high-redshift galaxy observations by providing unprecedented access to high-resolution rest frame optical imaging of reionization-era galaxies. Several of the first \textit{JWST} observations have indicated brighter galaxies at earlier times than most theoretical models predicted \citep{Finkelstein2023, Harikane2023c, Harikane2023a, Robertson2023}, and above extrapolations of data taken from the \textit{Hubble Space Telescope} \citep[\textit{HST};][]{Finkelstein2022}. Several theoretical studies have sought to explain this discrepancy by considering modifications to galaxy formation models, and improved treatment of dust effects \citep{Kannan2023, Shen2023}, different early stellar populations \citep{Steinhardt2023}, skewed mass-to-light ratios for early galaxies \citep{Inayoshi2022}, and , alternative dark energy models in the early universe \citep{Smith2022a}. In addition to findings of more faint active galactic nuclei (AGN) than expected \citep{Bischetti2022, Harikane2023b}, these early results hint at potential paradigm shifts in the role of various populations of galaxies in contributing to reionization mechanisms. Spectroscopic studies facilitated by the Near Infrared Spectrograph (NIRSpec) aboard the \textit{JWST} have also inferred constraints on ionized bubbles sizes around high redshift galaxies. Typically, these are based on damping wing absorption redward of the Ly$\alpha$ line \citep{Fujimoto2023, Hayes2023, Hsiao2023, Jung2023, Umeda2023}, but there have also been studies which use the transmission of the Ly$\alpha$ emission line itself to constrain the ionized bubble sizes around galaxies \citep{Saxena2023, Witstok2024}.

In parallel to galaxy observations, 21\,cm radio interferometers including the Low Frequency Array \citep[LOFAR;][]{vanHaarlem2013}, Hydrogen Epoch of Reionization Array \citep[HERA;][]{DeBoer2017, HERA2023}, Square Kilometer Array \citep[SKA;][]{Mellema2013}, and others are beginning to map the neutral hydrogen in the Universe. The complementary advantage of the 21\,cm observations is the ability to directly probe the IGM through line emission from the forbidden spin-flip hyperfine transition of neutral hydrogen. By looking for the redshifted signal from the EoR, these instruments will allow us to see the distribution of neutral and ionized gas in the IGM. These measurements will be critical for studying the ionized bubbles during the EoR as well as constraining cosmological parameters \citep{McQuinn2006, Mesinger2011, Liu2016, Park2019, Kannan2022b}.

In conjunction with the recent and upcoming observational capabilities, there has also been strong progress on developing theoretical methodologies to explore and interpret signals from the EoR. This is challenging because predictions of ionized bubble sizes will depend on the volume of the simulated box, which must be large enough to obtain a converged bubble size distribution. A variety of methods have been employed to study the processes of hydrogen reionization including perturbative evolution of density fields with ionized regions classified with excursion-set theory, furnishing rapid insights into ionized region classification and model parameter spaces \citep{Furlanetto2004a, Mesinger2011, Park2019, Fialkov2020, Munoz2020, Lu2024}. Beyond this, postprocessing of simulated dark matter halos with galaxy formation and radiative transfer models yields complementary insights about the physics of ionizing sources \citep{Ciardi2003, Iliev2007, McQuinn2009} and semi-analytic models of galaxy formation which include a variety of approximate or efficient prescriptions for including inhomogeneous reionization on the fly \citep{Mutch2016, Davies2023, Puchwein2023}. Finally, the most accurate but also computationally demanding are self-consistent radiation-hydrodynamic simulations, which couple the intricacies of galaxy formation and fully time-dependent radiative transfer physics \citep{Xu2013, Gnedin2014, Rosdahl2018, Pawlik2017, Lewis2022}.

In anticipation of the forthcoming 21\,cm data, numerous theoretical studies have sought to characterize the properties of ionized bubbles and their connections to the dominant processes in cosmic reionization. Different reionization schemes and source models can lead to different bubble morphologies, and the distribution of bubble sizes has emerged as a distinguishing feature between different reionization models \citep{McQuinn2007b, Majumdar2016}. Recent investigations indicate that quasars provide a negligible contribution to the ionizing radiation needed to reionize the Universe \citep{Eide2020, Jiang2022}, suggesting that stellar matter within galaxies are likely primarily responsible for hydrogen reionization. Discerning whether this process is driven by smaller numbers of highly luminous galaxies or the collection of numerous faint galaxies remains a critical goal of reionization research \citep{Bera2023, Yeh2023, Kostyuk2023}. The specific topological properties of the spread of ionized regions with potential to distinguish otherwise degenerate models is also a topic of current research. There has been a focus to determine the conditions for cosmological regions to ionize via an ``inside-out'' scenario, where the ionizing radiation originates from within the galaxy, or an ``outside-in'' scenario, where the ionizing radiation emanates from external sources and propagates into the galaxy to reionize the gas \citep{Choudhury2009}.

Topological analyses of the ionization field, redshift of reionization, and bubble morphologies have been fruitful in characterizing the distinctive signatures of reionization \citep{Friedrich2011, Busch2020, Thelie2022, Cain2023, Elbers2023}. There are a variety of ways to characterize the topology and morphology of reionization, including the utilisation of Betti numbers \citep{Giri2021, Kapahtia2021}, genus curves \citep{Lee2008}, and various alternative metrics \citep{Giri2020JOSS}. Other works focus specifically on forecasts for the 21\,cm signal \citep{Furlanetto2004b, Zaldarriaga2004}, or observed spectra \citep{Gazagnes2021}. 

A variety of methods exist for the detection and analysis of ionized bubble sizes within hydrodynamical simulations and semianalytic codes. Prominent among these are: the mean-free path method (MFP), which calculates bubble sizes by tracing rays from ionized cells to their first neutral cell encounter \citep{Mesinger2007}; spherical averaging (SPA), which calculates the largest ionized sphere based on a set ionization threshold \citep{Zahn2007}; friends of friends (FOF), which links ionized cells within a given distance to form a bubble \citep{Ivezic2014}, granulometry uses a 'sieving' process to count objects fitting within a given structure \citep{Kakiichi2017}; and the watershed method identifies bubbles by filling 'catchment basins' from local minima until neighboring basins intersect \citep{Lin2016}.

Many of these analyses have been performed using semi-numerical schemes that do not self-consistently model the impact of galaxies on hydrogen reionization. To build upon these existing studies, we perform an analysis of ionized bubble sizes and connections to their local cosmic environments within the \thesan simulations \citep{Garaldi2022, Kannan2022a, Smith2022, Garaldi2023}, which combines the galaxy formation model of IllustrisTNG \citep{Weinberger2017, Pillepich2018a, Pillepich2018b} with radiative process modeling including radiative transport \citep[\areport; ][]{Kannan2019}, non-equilibrium heating and cooling, and realistic ionizing sources throughout a large box of 95.5\,cMpc per side. The \thesan simulations have been used to make a wide range of EoR predictions \citep[e.g.][]{Kannan2022b, Qin2022, Borrow2023, Kannan2023, Yeh2023, Xu2023, Shen2024a, Shen2024b}.

The paper is organized as follows. In Section \ref{sec:methods}, we describe our methods, including a brief summary of the \thesan suite of simulations, the mean-free path (MFP) methodology, and our particular implementation of the MFP bubble size calculator. In Section \ref{sec:evolution}, we present our findings on the time evolution of bubble sizes, followed by an exploration of the impact of environmental factors on these sizes in Section \ref{sec:environment}. In Section \ref{sec:zreion}, we focus on the utility of the redshift of reionization as a probe of bubble sizes. Finally, we present our overall conclusions in Section \ref{sec:conclusion}. Supplementary discussions on grid resolution, smoothing scale, and simulation convergence are provided in Appendices \ref{appx:res}, \ref{appx:smooth}, and \ref{appx:conv}, respectively. All further mentions of ``reionization'' refer specifically to hydrogen reionization.

\begin{figure*}
    \centering
    \includegraphics[width=\textwidth]{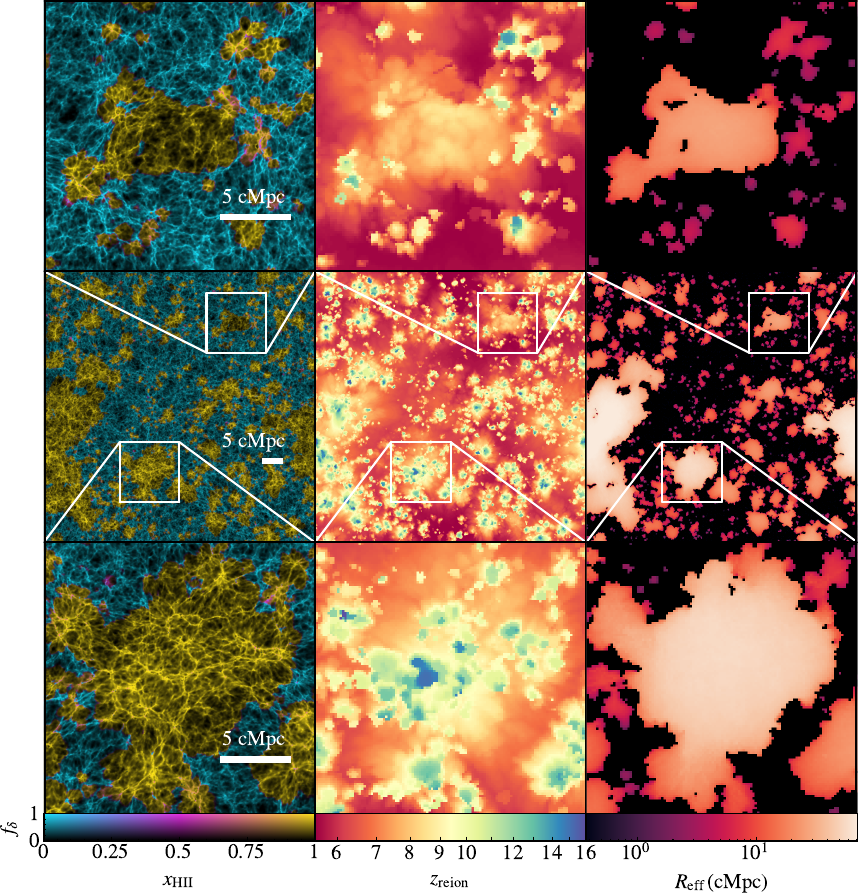}
    \caption{Image of the \thesanone box (centre row) and zoomed-in regions (top and bottom rows) illustrating the ionized fraction $x_{\HII}$ shaded by density $\rho$, redshift of reionization $z_\text{reion}$, and effective bubble size $R_\text{eff}$. The leftmost column shows a high-resolution projection through the central 10 per cent of the box with a two-dimensional colour bar such that the blue and yellow colours correspond to neutral and ionized gas, while the brightness scale corresponds to the logarithmic fractional overdensity, $f_{\delta}$. The middle column shows an equivalent slice through the intermediate resolution render box of the redshift of reionization, defined as the last redshift at which each cell is neutral ($x_{\HII} < 0.5$). The rightmost column shows the effective bubble size as seen by each cell, displaying sharp edges following the open or closed topology of the ionized regions. The cosmic web is not as prominent in the $z_\text{reion}$ and $R_\text{eff}$ images as in the density projection due to volume-weighting $x_{\HII}$ over scales larger than individual filaments widths ($L_\text{box}/512 \approx 0.2\,\text{cMpc}$).}
    \label{fig:HII_zreion_Reff}
\end{figure*}

\section{Simulation and analysis methods}
\label{sec:methods}
We briefly describe the \thesan reionization simulations in Section~\ref{sec:thesan}, focusing primarily on the ionization field as it pertains to this study. We then provide additional details about our procedure for determining bubble sizes using the MFP method in \ref{sec:mfp_method}, and our MFP implementation and data products in \ref{sec:mfp_data}.

\subsection{\thesan simulations}
\label{sec:thesan}
The \thesan project \citep{Kannan2022a, Garaldi2022, Smith2022, Garaldi2023} is a suite of large-volume ($L_{\text{box}} = 95.5~\text{cMpc}$) radiation-magneto-hydrodynamic (RMHD) simulations that simultaneously resolve large-scale structure for reionization and realistic galaxy formation using the IllustrisTNG model \citep{Pillepich2018a}, which is an updated version of the model used in Illustris \citep{Vogelsberger2013, Vogelsberger2014b, Vogelsberger2014a}. \thesan provides state-of-the-art resolution and physics for a reionization simulation of this volume and presents a unique opportunity to investigate connections between the topology of reionization and the galaxies responsible for reionizing the Universe. Photons from sources including stars and active galactic nuclei are tracked self-consistently in three energy bins (13.6 eV - 24.6 eV, 24.6 eV - 54.4 eV, and $\geq$ 54.4 eV). Stellar population properties including luminosities and spectral energy densities are determined by using the Binary Population and Spectral Synthesis library \citep[BPASS;][]{Eldridge2017}. \mbox{\textsc{thesan}} also incorporates non-equlibrium thermochemistry for tracking cooling by hydrogen and helium, as well as equilibrium cooling by metals. The high-resolution, fiducial simulation, \thesanone, has dark matter and baryonic mass resolutions of $3.1 \times 10^6\, \rm \Msun$ and $5.8 \times 10^5\, \Msun$, respectively. Haloes are resolved down to masses of $M_{\rm halo} \sim 10^8\, h^{-1}\, \Msun$ \citep{Garaldi2023}. The simulations use the efficient quasi-Lagrangian code \areport \citep{Kannan2019}, an extension of the moving mesh code \mbox{\textsc{arepo}}\xspace \citep{Springel2010, Weinberger2020} that includes radiative transport, to solve the fluid dynamics equations on an unstructured Voronoi mesh produced by approximately following the flow of the gas. The radiative transport equations are solved using a moment-based approach assuming the M1 closure condition \citep{Levermore1984, Dubroca1999}. The gravity solver uses the hybrid Tree-PM method, which estimates short-range gravitational forces using a hierarchical oct-tree algorithm \citep{BarnesHut1986}. Long-range gravitational potentials are calculated solving the Poisson equation using the Fourier method.

The \thesan simulations output 81 snapshots of particle positions and properties spanning redshifts from 20 to 5.5. These particle snapshots are converted into Cartesian grids with varying resolutions: 128, 256, 512, and 1024 cells per side. These Cartesian grid representations encapsulate volume-weighted properties, including the ionized fraction. In this paper, we use the renders with 512 cells per side. A discussion of convergence across different grid resolutions is provided in Appendix~\ref{appx:res}, but we note that we demonstrate satisfactory convergence of bubble size statistics. The resolution effects are minimal between global neutral fractions of 0.1 and 0.9, i.e. $\sim 10$ ($\sim 30$) per cent agreement for 256 (128) cells per side compared to the fiducial 512 resolution. The deviation is more significant at the very beginning and end of reionization, which is expected due to the increased importance of small-scale structures during these phases.

Large-scale environmental properties such as the dark matter overdensity, $\delta \equiv (\rho - \bar{\rho}) / \bar{\rho}$ where $\bar{\rho}$ is the mean density, are most meaningful after smoothing to a given filter scale. Bubble statistics are sensitive to small-scale features, so it is preferable to smooth out potentially transient or local features. Therefore, we also construct smoothed versions of the Cartesian outputs. Specifically, the particles are first binned at $1024^3$ resolution and then a periodic mass-conserving Gaussian smoothing kernel is applied with smoothing scales defined by standard deviations of 0\,ckpc (no smoothing), 125\,ckpc, 250\,ckpc, 500\,ckpc, and 1\,cMpc. We emphasize that lower resolution grids are coarse-grained versions of the high-resolution one with volume and mass weights propagated correctly. Results for bubble size evolution and comparison to the redshift of reionization are reported using the unsmoothed outputs, whereas the environmental analysis is performed using a smoothing scale of 125\,ckpc chosen to be above the grid resolution and extend up to typical galaxy separations such that the bubble sizes reflect the sources and environmental properties. The effects of the varying smoothing scales are discussed in Appendix~\ref{appx:smooth}, including potential sensitivities to over-smoothing of the ionization field. As the degree of smoothing increases, so does the median bubble size at each instance. This correlation is an expected outcome of the blurring of ionized regions induced by smoothing. The smallest bubbles tend to be approximately the same size as the smoothing scale (or the cell size in the unsmoothed case). The influence of smoothing scales on bubble sizes diminishes as reionization progresses, leading to substantial agreement once the majority of the gas becomes ionized and bubble sizes significantly exceed the smoothing scales.

\subsection{Mean-free path methodology}
\label{sec:mfp_method}
In this paper, we exclusively employ the MFP method for determining characteristic bubble sizes. We have chosen the MFP method because of its close correspondence with physical parameters, such as the mean-free path of photons emitted from the central cell. \citet{Lin2016} assert that both the MFP and watershed methods stand out as the most meaningful metrics for determining bubble sizes. Notably, the MFP determination of bubble sizes yields size distributions which are not biased toward artificially large or small bubble sizes in well-defined test cases \citep{Lin2016}, signifying that the peak of the bubble size distribution (BSD) aligns with the ``correct'' radius for a binary field with a single spherical ionized region. Unlike the watershed approach, the MFP method allows us to assign each cell a characteristic bubble size according to the ionization states of the surrounding gas. This corresponds to the effective bubble size ``seen'' by a region in the simulation. For example, if a small bubble begins to merge with a larger one, methods such as FOF and watershed might classify all the ionized cells as constituents of one extensive bubble with identical size characteristics. In contrast, the MFP method would retain more information about the progenitor bubbles because the MFP bubble size would be smaller for cells within the smaller bubble compared to the larger one. This distinction is useful for studying environmental effects on bubble sizes, as well as in drawing comparisons with photon mean-free path and line-of-sight results.

In our analysis, we identify bubble sizes by extending 192 rays in isotropically-determined directions, following the HEALPix prescription of equal area segmentations of a sphere \citep{Gorski1999}. This ensures uniform sampling of the directions of the rays. The number of HEALPix rays was chosen to account for the complex bubble morphology, while keeping computation time low. We find less than 5\% deviation in median bubble sizes compared to using 768 rays per cell (and variations $\lesssim 10\%$ compared to 48 HEALPix rays). We calculate the length of each ray between the starting point and the first instance that it intersects a cell possessing an ionization fraction below a threshold value of 0.5. We utilize second-order ray tracing with gradient limiters to maintain values in the range $[0,1]$ and avoid overshooting neighbouring values. For any neutral cells where the ionized fraction is $x_{\HII} < 0.5$, the MFP bubble size is assigned a value of zero. This choice of threshold, $x_{\HII} = 0.5$ is arbitrary, and median bubble sizes differ by $\lesssim 10\%$ compared to thresholds of $x_{\HII} = 0.1$ and 0.9, for most of the EoR. There are larger relative variations ($\lesssim 30 \%$) at high neutral fraction ($x_\HI \gtrsim 0.8$), when bubbles are small. The algorithm we employ is similar to the method utilized by Tools21cm \citep{Giri2020JOSS}, specifically adapted to account for periodic boundary conditions and systematic sampling of every ionized cell with uniformly distributed ray directions according to the HEALPix prescription. Each ray is followed through the ionization field which has been linearly interpolated between neighboring cells, again with second-order ray tracing. The ray's length is recorded once it surpasses the neutral threshold value. We allow a maximum ray length of twice the box size (191\,cMpc) due to the periodicity of the grid.

\begin{figure}
    \centering
    \includegraphics[width=\columnwidth]{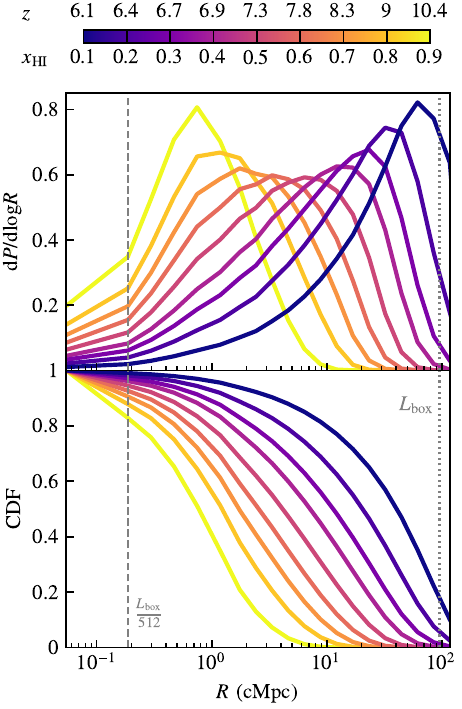}
    \caption{Ionized bubble size distributions for the fiducial \thesanone simulation. Each line shows the distribution at a different neutral fraction, or equivalently, redshift. All statistics are volume-weighted. The top panel shows the probability distribution function for bubble sizes and the bottom panel quantifies the cumulative distribution function for bubbles with effective radii $\geq R$. Bubble sizes grow over time as reionization progresses with the formation and growth of bubbles that eventually percolate to create much larger interconnected bubbles. There is a sharp peak at $R \sim 1\,\text{cMpc}$ for $x_{\HI} = 0.9$ (yellow line) as most bubbles have recently formed and have not yet merged. There is another sharp peak at $R \gtrsim 50\,\text{cMpc}$ for $x_{\HI} = 0.1$ (dark purple line) once most of the ionized cells are part of large regions spanning the simulation volume.}
    \label{fig:unsmoothed_bsd}
\end{figure}

\begin{figure}
    \centering
    \includegraphics[width=\columnwidth]{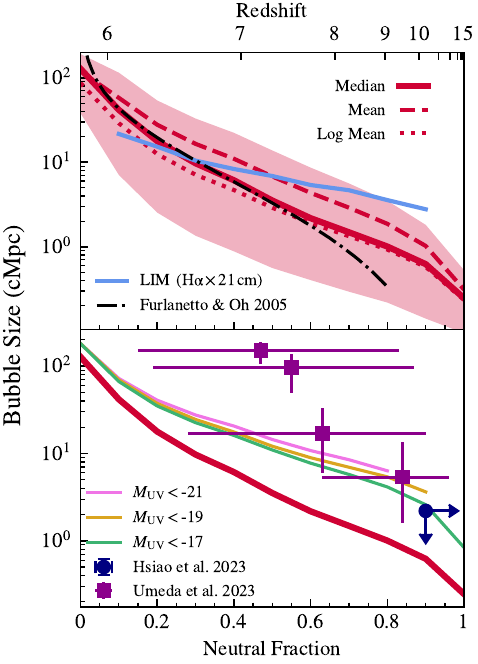}
    \caption{Redshift and global neutral fraction evolution of the characteristic bubble sizes for the fiducial \thesanone high-resolution simulation renderings. The median, mean, and log mean bubble sizes are shown as a function of neutral fraction and redshift in the top panel. The shaded region indicates the $16^\text{th}$ to $84^\text{th}$ percentile range. We compare to the \thesan line intensity mapping (LIM) results from \citet{Kannan2022b} in light blue, which are fairly consistent at redshifts below about 9. We also compare with the analytic model from \citet{Furlanetto2005} in the black dash-dotted line. The bottom panel shows median bubble sizes weighted by the number of bright haloes according to the threshold dust-corrected UV magnitude $M_{1500}$. These are compared with the observationally inferred bubble size constraints from \citet{Hsiao2023} in the dark blue circle and from \citet{Umeda2023} in the purple squares.}
    \label{fig:mfp_bubble_size}
\end{figure}

\subsection{Mean-free path data products}
\label{sec:mfp_data}
We apply the MFP method to the Cartesian outputs with a resolution of 512 cells per side such that the cell sizes are $L_\text{cell} = 186.5$\,ckpc. We compute the \textit{effective} MFP bubble size, denoted by $R_\text{eff}$, for each cell by averaging multiple second-order ray-traces along 192 HEALPix directions. Positioning these bubbles as ``centred'' on the cell itself permits $R_\text{eff}$ to be considered as a distinct property of each cell, facilitating direct comparisons with other spatial attributes. Note that we consider $R_{\rm eff}$ to be the average of the lengths of the radial rays, not as the effective radius of the volume sampled by the 192 rays. These statistics are, by construction, volume-weighted as each equal-volume Cartesian grid cell gives an equal contribution to population statistics (e.g. BSDs and median bubble sizes).

Additionally, we generate a BSD histogram encompassing all ray lengths originating from ionized cells throughout the simulation box. For the smoothed renders we also compile two-dimensional histograms of ray lengths and the local dark matter overdensity $\delta$ and other environmental quantities of the starting cell to explore environmental effects. The global BSD histogram is saved with a bin resolution of $L_\text{cell}/8$, representing a characteristic scale on which the ionized fraction may change. For the unsmoothed case, we perform the MFP analysis on all 81 of the \thesan snapshots to achieve maximum redshift resolution. For the smoothed outputs, we report results from the MFP bubble size analysis for nine snapshots, chosen such that the global volume-weighted neutral fraction increases from $x_{\HI} = 0.1$ to $0.9$ in increments of 0.1.

In Figure~\ref{fig:HII_zreion_Reff}, we illustrate the density $\rho$ coloured by the ionized fraction $x_{\HII}$, redshift of reionization $z_\text{reion}$, and effective bubble size $R_\text{eff}$ in the left, middle, and right columns, respectively. The snapshot corresponds to a redshift when the global neutral fraction is one half. The middle row displays the entire \thesan box ($L_\text{box} = 95.5$\,cMpc), while the top and bottom rows offer detailed views of two zoomed-in regions. The leftmost column shows neutral gas in blue and ionized gas in yellow, with variations in colour intensity denoting the density, with lighter (darker) shades corresponding to more (less) dense regions. The left column is a high-resolution ($2048^2$) projection through 10 per cent of the box, whereas the middle and right columns show the equivalent slice through the 512 cells per side render box. This figure illustrates the complexity of interconnected ionized regions along with the richness of the large-scale structure. The redshift of reionization retains the structure of the bubble evolution and the instantaneous bubble size ($R_{\rm eff}$) shows the morphology at a neutral fraction of 0.5.

\begin{figure}
    \centering
    \includegraphics[width=\columnwidth]{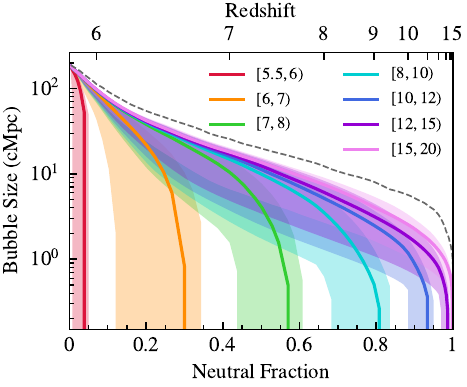}
    \caption{Bubble size trajectories over time grouped by the local redshift of reionization \zreion for the high-resolution fiducial \thesanone model. Each solid curve shows the median effective bubble size $R_\text{eff}$ from all histories within each \zreion range, while the shaded regions are the $16^\text{th}$ to $84^\text{th}$ percentile ranges. The dashed grey curve shows the largest bubble size measured from each snapshot. We define \zreion of each cell to be the lowest redshift at which the cell crossed from neutral ($x_{\HII} < 0.5$) to ionized ($x_{\HII} \geq 0.5$). By $z \sim 6$, regions are flash-ionized because they immediately join a large pre-existing bubble.}
    \label{fig:trajectories}
\end{figure}

\section{Evolution of bubble sizes}
\label{sec:evolution}
As the Universe becomes reionized by the emission of ionizing radiation from astrophysical sources, the ionized gas bubbles grow and merge in a process called percolation. As neighbouring regions of ionized gas run into one another, percolation causes sudden and dramatic increases in bubble sizes within a short period of time. The top panel of Figure~\ref{fig:unsmoothed_bsd} illustrates the BSDs with each line showing the distribution for neutral fractions between 0.1 and 0.9 at select redshifts between 6 and 11, while the bottom panel quantifies the cumulative distribution functions reflecting the probability that an ionized bubble has an effective radius above the given $R$). The distributions are volume-weighted as each Cartesian grid cell contributes an equal number of MFP measurements. The bubbles tend to grow in size throughout reionization with the distributions being unimodal and most strongly peaked during the early ($x_{\HI} \gtrsim 0.9$, yellow line) and late ($x_{\HI} \lesssim 0.1$, dark purple line) phases of the EoR. In the former case, individual bubbles are forming and growing in isolation, yielding a fairly consistent characteristic size. Conversely, toward the end of the EoR, most of the ionized gas resides within extensive bubbles that have formed through the percolation of many smaller bubbles. The unimodality of the distributions indicates that there continues to be a well-defined characteristic bubble size throughout reionization, despite the complex growth and percolation processes. In addition, since we are considering volume-weighted statistics, the largest bubble will be primarily responsible for the distributions on the largest scales, while contributions from smaller bubbles are systematically repressed.

At higher redshifts when the first ionized bubbles are beginning to grow around ionizing sources, these structures tend to be small, on scales of a few hundred comoving kiloparsecs (ckpc). By a redshift of $z \sim 9$, the characteristic bubble size surpasses the threshold of $R_\text{eff} = 1$\,cMpc. Figure~\ref{fig:mfp_bubble_size} shows the median, mean, and log mean bubble sizes over time, with the shaded region representing the $16^\text{th}$ to $84^\text{th}$ percentile range in the top panel. The global neutral fraction is plotted on the lower horizontal axis, while corresponding redshifts are marked on the upper horizontal axis. Predicted line-intensity mapping results from \citet{Kannan2022b} are shown in light blue and are consistent with the MFP estimates of bubble sizes found here, especially at lower redshifts. Additionally, the analytic model assuming no bubble overlap from \citet{Furlanetto2005}, shown as the black dash-dotted line, also agrees well with our bubble size measurements. 

In the bottom panel of Figure~\ref{fig:mfp_bubble_size}, we show median bubble sizes weighted by the number of haloes brighter than the threshold dust-attenuated UV magnitude $M_\text{UV}$ according to the dust correction from \citet{Gnedin2014} and \citet{Vogelsberger2020}. Since brighter sources tend to reside in larger bubbles, these curves lie above the overall bubble sizes, more closely reflecting observations that are biased toward more luminous galaxies. Bubble size constraints from Ly$\alpha$ damping wings found in high-$z$ galaxy spectra from \textit{JWST} are shown with the dark blue circle \citep{Hsiao2023} and purple squares \citep{Umeda2023}. The bubble size ($R < 0.2\, \rm pMpc$) and neutral fraction ($x_{\HI} > 0.9$) inferred by \citet{Hsiao2023} for the $z = 10.17$ triply-lensed galaxy MACS0647--JD are consistent with our findings. However, the bubble sizes reported in \citet{Umeda2023} are higher than our estimates, particularly at lower redshifts toward the end of the EoR. This discrepancy is possibly due to their assumption of a homogeneous IGM, potentially leading to a systematic overestimation of neutral fractions during the EoR. Additionally, \citet{Keating2023a} have found that many of these Ly$\alpha$ damping wing observations can be explained with smaller bubble sizes and larger intrinsic Ly$\alpha$ fluxes from the host galaxies themselves. Late-time BSDs are also limited by the box size, which could be addressed by running larger-volume simulations in the future within the same framework. As a more concrete next step, we plan to investigate analytic biases with end-to-end forward modeling for a more fair comparison. Finally, with only a few observations it is difficult to simultaneously constrain the local line-of-sight bubble size and global evolution of the neutral fraction.

\begin{figure}
    \centering
    \includegraphics[width=\columnwidth]{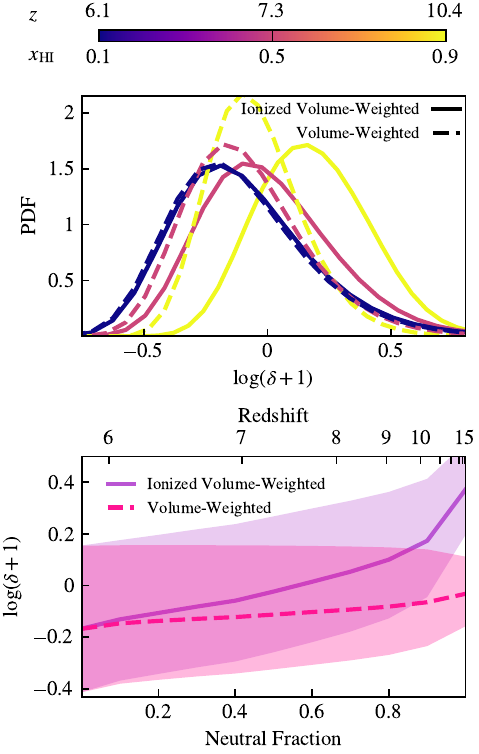}
    \caption{\emph{Top panel}: Probability density functions of local environmental overdensity for the entire simulation volume (dashed lines) and just the ionized volume (solid lines), shown at times when the global neutral fraction is 0.1, 0.5, and 0.9. \emph{Bottom panel}: The median overdensity and $16^\text{th}$ to $84^\text{th}$ percentile range as functions of neutral fraction and redshift. The distributions are nearly non-overlapping at the earliest times but quickly converge as the entire simulation box becomes ionized.}
    \label{fig:delta_pdfs}
\end{figure}
\begin{figure*}
    \centering
    \includegraphics[width=\textwidth]{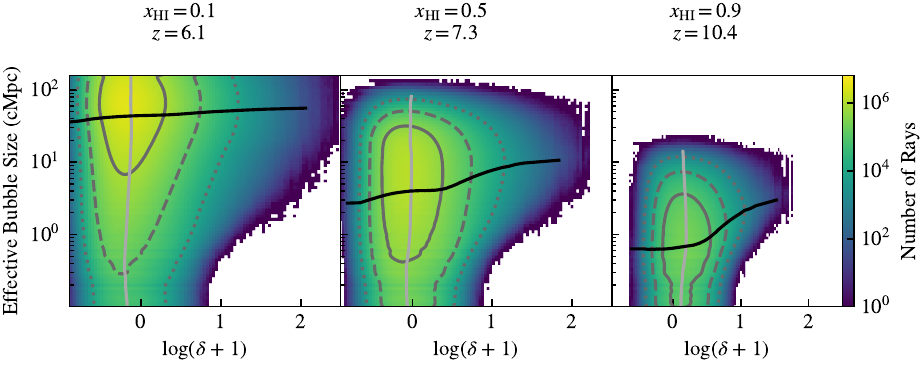}
    \caption{Characteristic ionized bubble sizes compared to overdensity for \thesan high resolution cube smoothed on 125ckpc scales at different neutral fractions $x_{\HI}$ and their corresponding redshifts. Grey contours show the 1, 2, and 3 $\sigma$ regions in solid, dashed, and dotted lines, respectively. While the contours do not indicate a strong correlation between overdensity and bubble sizes, the shape of the histogram indicates that regions with high local overdensities preferentially have large bubbles. The black solid lines show the median bubble size at each overdensity, and the light grey lines show the median overdensity at each bubble size.}
    \label{fig:env_2d}
\end{figure*}

The variation in bubble size is relatively small at the beginning and end of the EoR. At very high redshifts ($z \gtrsim 10$), ionized bubbles have only recently begun to form around luminous objects and have not yet grown enough to instigate percolation, a process that rapidly increases individual bubble sizes. Consequently, nearly all of the first ionized bubbles remain below 1\,cMpc in size, as most galaxies are not yet massive enough to sustain high star-formation rates. As the bubbles continue to expand due to radiation continuously ionizing more and more gas around the source, the bubbles begin to percolate and coalesce. This leads to a relatively large spread in bubble sizes during the intermediate stages of the EoR mirroring the asynchronous nature of reionization ($6 \lesssim z \lesssim 10$). At even later stages ($z \lesssim 6$), as the majority of bubbles become encompassed by larger neighbouring bubbles, the sizes reflect the vast network of interconnected ionized regions. By this point near the end of the EoR, these immense bubbles housing the majority of the ionized gas all merge with each other until there is effectively one large bubble filling nearly the entire simulation box, which is dominant in volume-weighted statistics. This is indicated by the median bubble size at neutral fractions close to zero becoming larger than the side length of the box.

Regions that become ionized at different times during the EoR have dramatically different size evolution patterns. To highlight this, we explore the typical bubble growth trajectories for regions reionized at different redshifts. Categorizing by the redshift of reionization, \zreion, allows us to identify different populations, each with unique bubble size histories over time. The \zreion of a simulation cell is defined as the lowest redshift at which the cell transitions from neutral ($x_{\HI} < 0.5$) to ionized ($x_{\HI} \geq 0.5$). We examine the evolution of bubble sizes for seven \zreion ranges delineated by the bin edge values: \zreion $\in [5.5, 6, 7, 8, 10, 12, 15, 20]$. The median bubble size trajectories for these different \zreion ranges are shown in Figure~\ref{fig:trajectories}.

Cells with \zreion $\geq$ 15 (pink line) are the first cells to become ionized and initiate the reionization process. The bubbles around these cells quickly reach an effective radius of about 1\,cMpc, then begin to grow roughly exponentially with the declining neutral fraction until they reach the maximum allowed bubble size of $\sim 200$\,cMpc, marking the point when the entire box is reionized. Cells with \zreion $\gtrsim$ 10 follow a similar trajectory, indicating that most of these cells become ionized without immediately joining a pre-existing large bubble. The bubble sizes for these early-reionized cells start small ($\sim$\,cMpc scales) and then exhibit a gradual growth pattern, similar to delayed versions of the first ionized regions.

Once \zreion $\lesssim$ 10, many of the newly ionized cells promptly merge with existing ionized bubbles, which have sizes ranging from a few cMpc to tens of cMpc for 7 $\lesssim$ \zreion $\lesssim$ 10 and 6 $\lesssim$ \zreion $\lesssim$ 7, respectively. Following the main mergers into large pre-existing ionized regions, the bubble sizes grow at a more subdued rate similar to the growth observed in earlier-reionized cells. This gradual assimilation process is coherent, considering that the later-ionized cells are likely integrated into the very same bubbles that the early-reionized cells began to form. Thus, this slower expansion may track a characteristic bubble growth rate persisting throughout the procession of reionization. The large vertical scatter in bubble size indicates that inside-out reionization is still occurring even at $z \lesssim 10$ \citep{Kannan2022a, Borrow2023}. New small bubbles are still forming in neutral areas even after much of the IGM has become ionized and percolation of large bubbles has become a significant cause of bubble size increases.

The most rapid growth rates are observed right at the culmination of reionization, specifically after a redshift of 6, shown in the red line in Figure~\ref{fig:trajectories}. These late-reionization regions are ``flash-ionized'' as they immediately become part of the last large pre-existing bubble which already fills most of the simulation box. These cells represent some of the final neutral island regions in the simulation. Thus, no clear bubble boundaries are discernible by the time they become ionized, as the ionized gas essentially forms one final space-filling bubble.

\begin{figure}
    \centering
    \includegraphics[width=\columnwidth]{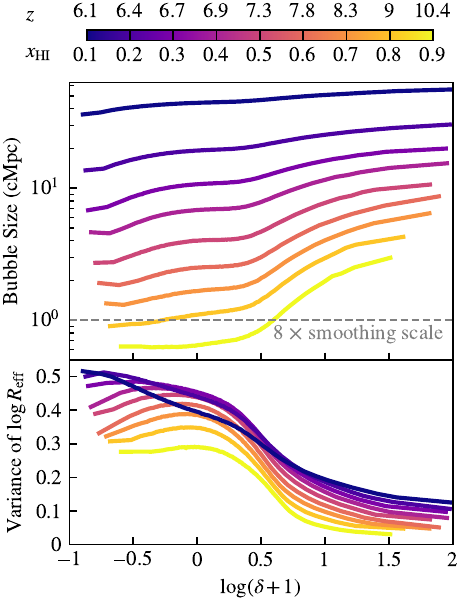}
    \caption{Median bubble size $R_\text{eff}$ as a function of overdensity at different ionized fractions for the fiducial model smoothed on 125\,ckpc scales. The grey dashed line shows 8 times the smoothing scale. Larger overdensities tend to have larger bubble sizes, especially at early times when the bubbles are initially forming. At later times, when most of the gas in the simulation box is ionized, bubble sizes have a much weaker dependence on overdensity, likely due to the prevalence of few large bubbles which encompass most of the ionized volume. Overall, there is larger variance in bubble sizes at lower overdensities. The variance flattens out with time such that void-like regions start to decrease while overdense regions increase, which in the latter case is primarily driven by the increasing number of bright objects with time.}
    \label{fig:size_vs_delta}
\end{figure}

\begin{figure*}
    \centering
    \includegraphics[width=\textwidth]{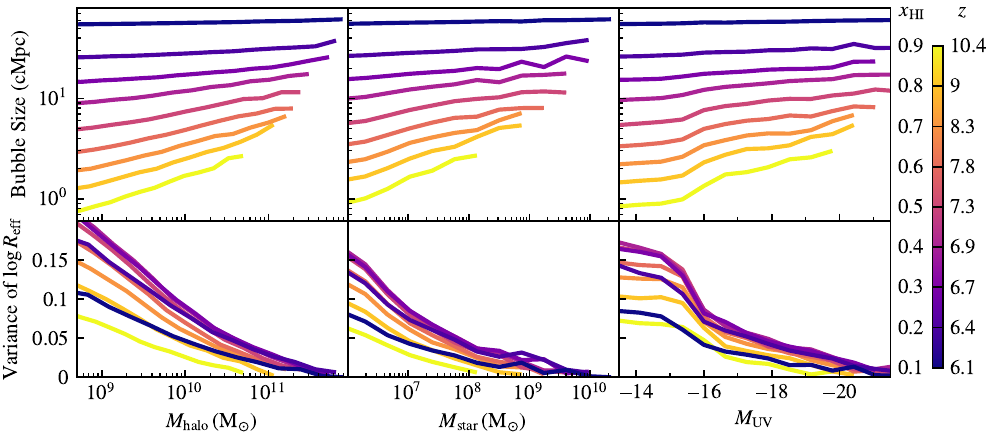}
    \caption{Similar to Figure~\ref{fig:size_vs_delta}, we compare median bubble sizes and variances as functions of galactic halo mass, stellar mass, and UV magnitude. Haloes are resolved down to masses of $M_{\rm halo} \sim 10^8\, h^{-1}\, \Msun$, so resolution effects should not be significant in the range considered. We find similar trends to those for overdensity but as these halos host the sources of reionization the dependence extends across the entire resolved halo range, exhibiting a characteristic power-law behaviour.}
    \label{fig:galaxies}
\end{figure*}

\section{Connections to the local environment}
\label{sec:environment}
In an effort to understand the conditions that influence ionized bubble properties, we investigate the relationship between the sizes of ionized bubbles and the local environmental overdensity. We smooth the overdensities from the Cartesian grid on the scale of 125 ckpc such that the overdensities are probing the local environment as opposed to smaller scale density fluctuations. We begin by examining the distribution of overdensities within our simulated volume at different stages of reionization. These distributions, both volume-weighted and ionized volume-weighted, are shown in the top panel of Figure~\ref{fig:delta_pdfs}. The ionized volume-weighted distributions are shifted toward higher overdensities early in reionization as compared to their full volume-weighted counterparts. This indicates that regions that are more overdense tend to be ionized early and that many of the first regions that are reionized reside in locally overdense environments. This is also reflected in the bottom panel of Figure~\ref{fig:delta_pdfs} which shows the median and $16^\text{th}$ to $84^\text{th}$ percentile range of overdensity with both weighting schemes. We again see that at high redshifts, the ionized volume-weighted median of overdensity is larger than the volume-weighted median. These medians and one sigma ranges converge to the same values by the end of reionization when nearly all of the volume of the box contains ionized gas. Specifically at global neutral fractions of $x_{\HI} = \{0.9, 0.5, 0.1\}$ the bias in the median overdensity is $\log ((\delta_\text{\HII}+1)/(\delta+1)) = \{0.24, 0.09, 0.02\}$.

We move on to investigate the specific correlation between local environmental overdensity and ionized bubble sizes. For consistency with the overdensity analysis, we smooth the bubble sizes on the same 125 ckpc scales to focus on the most ``significant'' bubbles in the local area. In Figure~\ref{fig:env_2d} we show two-dimensional histograms in the $\delta$--$R_\text{eff}$ plane highlighting various environmental effects, where each panel corresponds to a different neutral fraction and its corresponding redshift. The contours outline the central 68, 95, and 99.7 percentile ranges to guide the eye in discerning the volume-weighted statistics. The solid black curves show the median effective bubble size for each of the overdensity bins, and the light grey curves show the median overdensity for each of the bubble size bins. In summary, the local overdensity has the strongest impact early on during the formation phases. As reionization progresses there is a significant deficit of small bubbles in overdense regions whereas underdense regions can still harbour neutral structures.

At early times during the EoR, when 90 per cent of the gas remains neutral, bubble sizes are generally small with $R_\text{eff} \lesssim$ a few cMpc, but there is a slight correlation between overdensity and bubble sizes, as shown by the asymmetrical shape of the 2D histograms in Figure~\ref{fig:env_2d}. By examining the central 68 percentile contour in the rightmost panel of Figure~\ref{fig:env_2d}, we see that the smallest bubbles primarily reside in regions with $0 \lesssim \log (\delta + 1) \lesssim 0.5$. This indicates that initially these bubbles are primarily forming in overdense regions before expanding into surrounding areas with a wider range of densities, as depicted by the broader shape of the contours in the higher bubble size region of the panel. As reionization progresses and more of the gas becomes ionized, the overdensity dependence washes out, likely due to ionized regions joining pre-existing bubbles. By the time the EoR is almost over and 90 per cent of the gas is ionized, most of the bubbles are large ($R_\text{eff} \gtrsim 10\, \rm{cMpc}$) and there is no significant correlation with particular overdensity values.

We focus on the relationship between bubble sizes and overdensity to determine whether bubbles predominantly form and grow within high-density regions or low-density regions. Higher-density regions typically contain more ionizing sources, but their higher density of initially neutral gas makes them more difficult to ionize. In the top panel of Figure~\ref{fig:size_vs_delta} we show the median bubble size in relation to the corresponding overdensity at which it is centred. We employ adaptive binning in overdensity to improve the statistical significance at the highest and lowest densities. Across all redshifts that we examined, a clear pattern emerges where the bubble size increases with overdensity. This indicates that the largest bubbles are more likely to originate from and surround denser regions. However, this trend becomes less pronounced as reionization progresses and bubbles undergo percolation, coalescing into fewer ionized regions. As a result of structure formation the late-time ionized landscape is volume-biased towards being underdense.

In the bottom panel of Figure~\ref{fig:size_vs_delta} we also illustrate the relationship between the variance in logarithmic bubble sizes and overdensity. Notably, the variance is significantly higher within low-density regions compared to more overdense areas. This is likely due to void-like regions becoming ionized by pre-existing bubbles growing into them (outside-in reionization) as well as some sources creating ionized bubbles within the low-density regions themselves (inside-out reionization). Like the median bubble size, the variance in bubble sizes also flattens out over time as reionization progresses. The variance of bubble sizes in the low density regions starts to decrease at late times while the variance in overdense regions continues to increase throughout the EoR. The increase in variance in dense regions is likely due to the increasing number of bright sources over time. Although many of the largest bubbles originate from areas of higher overdensity, the formation of new sources in neutral regions results in the simultaneous creation of many smaller bubbles as well. This can explain the continual increase in bubble size variance in overdense regions as the largest bubbles get bigger, but the small bubbles continue to form due to the high density of haloes.

Further analysis reveals that, at $z=10$, bubbles situated in the highest overdensity regions are approximately 20 per cent larger than the global median size. However, this disparity diminishes to less than five per cent by $z=7$. This is again due to ionized bubbles primarily forming in very overdense regions early on and then expanding and merging to encompass underdense regions later in the EoR.

We also investigate how bubble sizes around galaxies depend on the galaxy properties, specifically halo mass stellar mass, and UV magnitude, as these relationships are of much interest to the observational and theoretical communities \citep{Davies2023, Hayes2023}. Our findings, summarized in Figure~\ref{fig:galaxies}, reveal that the median bubble size around different mass halos captures the correlation of these characteristics with expected bubble sizes throughout the EoR. Additionally, we examined the variance in bubble sizes for these galactic properties. The observed trends closely align with the local overdensity analysis: more massive, brighter galaxies are associated with larger bubble sizes with lower variance, especially during the earlier phases of reionization. As the EoR concludes, both the median and variance in bubble sizes flatten out, suggesting a weaker connection between galaxy properties and effective bubble sizes. This is consistent with the notion that bubbles have undergone percolation, resulting in a small number of dominant ionized regions.

\begin{figure}
    \centering
    \includegraphics[width=\columnwidth]{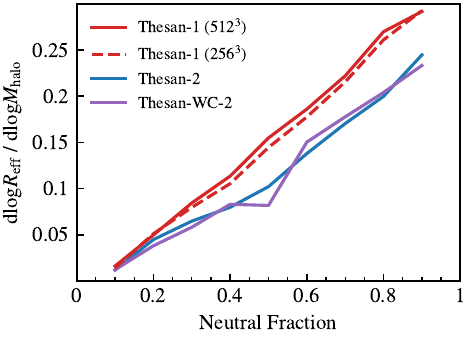}
    \caption{Power-law slope for $R_{\rm eff}$ vs $M_{\rm halo}$ over neutral fraction for the fiducial, high-resolution \thesanone run (red line) as well as the medium resolution runs \thesantwo (blue line) and \thesanwc (purple line). We also compare to \thesanone on a grid with 256 cells per side (red dashed line) for consistency with the \thesantwo grid resolution. For all runs, the power law is steepest (highest ${\rm d}R_{\rm eff} / {\rm d}M_{\rm halo}$) early in the reionization process driven by the cosmic structure formation but flattens as bubbles expand into void regions.}
    \label{fig:powerlaw}
\end{figure}

We quantify the degree of correlation between the effective bubble size and halo mass by fitting a power-law model to the $R_{\rm eff} \ {\rm vs.} \ M_{\rm halo}$ curves at each of the neutral fractions shown in Figure~\ref{fig:galaxies}. We note the best-fit power-law slope only considers halo masses in the range $M_\text{halo} \geq 10^9\, \Msun$. The best-fit parameter values are listed in Table~\ref{tab:pl}. The evolution of this metric throughout the EoR as shown in Figure~\ref{fig:powerlaw} clearly shows the strength of the early correlation and robustness of the flattening with time. We plot this power-law slope, $\text{d}\log R_{\rm eff}/\text{d}\log M_{\rm halo}$, as a function of the global neutral fraction for the fiducial, high-resolution \thesanone run (red curve) as well as the medium resolution runs \thesantwo and \thesanwc (blue and purple curves). Specifically, \thesantwo uses the same model as \thesanone but with two (eight) times coarser spatial (mass) resolution, and \thesanwc exhibits weak convergence of $x_{\HI}(z)$ by increasing the birth cloud escape fraction from 0.37 to 0.43 to compensate for additional unresolved low-mass haloes. Each of the \thesantwo models has been analyzed on a Cartesian grid with 256 cells per side, so we also show the small difference by re-calculating \thesanone on this lower resolution grid for consistency (red dashed curve). More details of the different models can be found in \citet{Kannan2022a} and \citet{Garaldi2023}. Our analysis reveals that the higher-resolution \thesanone simulation exhibits the steepest power-law slope between bubble size and halo mass. We attribute this to the abundant low-mass haloes and dense structures of clumps and filaments that are less resolved in the lower-resolution runs. Further discussion of the convergence of bubble size statistics across the different resolution \thesan runs is contained in Appendix~\ref{appx:conv}.

\begin{table}
    \centering
    \caption{Power-law fits to the bubble size-halo mass relationship (shown in the top left panel of Figure~\ref{fig:galaxies}) in the form $R_{\rm eff} / {\rm cMpc}= R_0 \, (M_{\rm halo} / \Msun)^\alpha$ for each neutral fraction, $x_\HI$ (and corresponding redshift, $z$) for \mbox{\textsc{thesan-1}}. The power-law slope parameter, $\alpha$, is plotted in Figure~\ref{fig:powerlaw} along with those from lower resolution simulations.}
    \addtolength{\tabcolsep}{-2.75pt}
    \renewcommand{\arraystretch}{1.1}
    \begin{tabular}{l|ccccccccc}
        \hline
        $x_\HI$\;\; & 0.1 & 0.2 & 0.3 & 0.4 & 0.5 & 0.6 & 0.7 & 0.8 & 0.9 \\
        $z$ & 6.1 & 6.4 & 6.7 & 6.9 & 7.3 & 7.8 & 8.3 & 9.0 & 10.4 \\
        \hline
        $\alpha$ & 0.016 & 0.050 & 0.084 & 0.11 & 0.15 & 0.19 & 0.22 & 0.27 & 0.29 \\
        $R_0$ & 39 & 9.0 & 2.5 & 0.87 & 0.21 & 0.067 & 0.021 & 0.0051 & 0.0020 \\
        \hline
    \end{tabular}
    \addtolength{\tabcolsep}{2.75pt}
    \renewcommand{\arraystretch}{0.9090909090909090909}
    \label{tab:pl}
\end{table}

\begin{figure}
    \centering
    \includegraphics[width=\columnwidth]{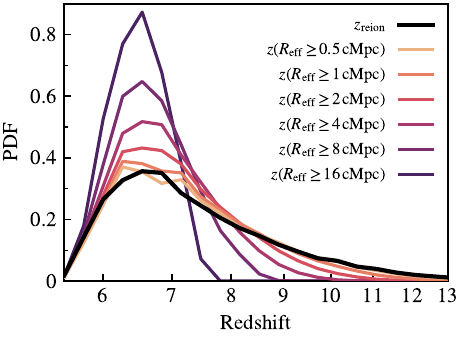}
    \caption{Probability density functions for \zreion (in black) and the redshift at which the bubble size surpasses various thresholds. We see that \zreion is a good tracer of bubble sizes below about 1\,cMpc, but time delays between reionization and bubble growth weaken the correspondence between \zreion and larger bubble size statistics.}
    \label{fig:z_reion_z_Reff_1d}
\end{figure}

\begin{figure}
    \centering
    \includegraphics[width=\columnwidth]{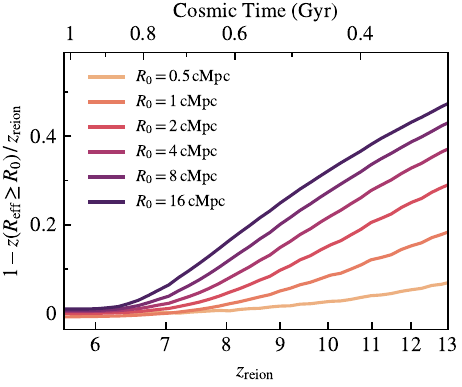}
    \caption{Relative difference between redshift of reionization and redshift at which the bubble size exceeds some threshold size $R_0$ plotted against \zreion. This illustrates the delay between reionization and bubble growth throughout the EoR, with the largest time lags corresponding to higher redshifts.}
    \label{fig:z_reion_z_Reff_difference}
\end{figure}

\begin{figure}
    \centering
    \includegraphics[width=\columnwidth]{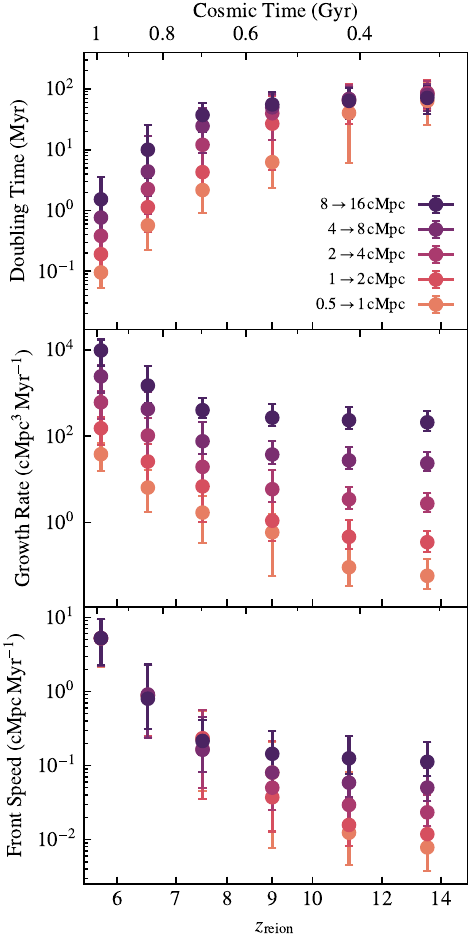}
    \caption{\emph{Top panel}: Median and $16^\text{th}$ to $84^\text{th}$ percentile range of doubling time grouped by local \zreion. \emph{Middle panel}: Median and $16^\text{th}$ to $84^\text{th}$ percentile range of growth rate of bubbles (change in effective volume over doubling time) in the same \zreion ranges. \emph{Bottom panel}: Median and $16^\text{th}$ to $84^\text{th}$ percentile range of the speed of ionization fronts in the same $z_{\rm reion}$ ranges, assuming spherical bubbles. Each colour corresponds to corresponding bubble size thresholds, e.g. dark purple points show the doubling times and growth rates for bubbles growing from 8\,cMpc to 16\,cMpc in radius. The later a region is locally reionized, the longer the doubling time, and the faster the growth rate. This is consistent with the picture of slow bubble growth early in reionization and ``flash-ionization'' and runaway percolation at lower redshifts.}
    \label{fig:doubling_time}
\end{figure}

\section{Connections to the Redshift of Reionization}
\label{sec:zreion}
The redshift of reionization, denoted as \zreion, serves as a powerful probe of the spatiotemporal distribution of ionizing radiation during the EoR. In this work, we define \zreion for each point in space as the latest redshift at which the Cartesian grid cell transitions from a neutral to an ionized state, specifically when the ionized hydrogen fraction reaches $x_{\HII} \geq 0.5$. To obtain an accurate estimate of \zreion, we employ linear interpolation between the simulation snapshots, each separated by a time cadence of $\sim10$\,Myr. 

To investigate the relationship between \zreion and $R_\text{eff}$, we calculate the last redshift at which each cell has an effective bubble size that crosses a specified threshold radius, denoted as $z(R_\text{eff} \geq R_0)$. As a first glance, we present the distributions of \zreion and $z(R_\text{eff} \geq R_0)$ for various threshold radii in Figure \ref{fig:z_reion_z_Reff_1d}. Our analysis reveals a reasonably good agreement between the distributions of \zreion and the redshift at which bubble sizes surpass a threshold radius when $R_\text{eff} \lesssim 1\,\text{cMpc}$.

As the threshold for bubble size increases, two notable shifts occur in the distribution: the peak moves slightly towards lower redshifts, and the high-redshift tail contracts. At high redshifts, bubbles are predominantly small, leading to the omission of many of the earliest ionized regions when higher bubble size thresholds are applied. This results in the attenuation of the slow-rising high-redshift tail observed in \zreion. Additionally, the distributions are slightly shifted towards lower redshifts compared to \zreion due to the time it takes between a cell becoming ionized and the subsequent growth of a bubble to the threshold size. The time delay between reionization and growing to bubble sizes larger than $\sim 1 \, \rm cMpc$, washes out the correspondence between \zreion and the redshift at which a bubble grows to that size.

In Figure~\ref{fig:z_reion_z_Reff_difference}, we more closely examine the time lags between cell reionization and the attainment of specific bubble sizes. We find that larger bubble size thresholds directly translate to longer time lags between these two events, while $z_{\rm reion}$ is a good tracer of small bubble sizes below approximately 1\,cMpc due to much shorter time lags. For all thresholds, this relative redshift difference with respect to \zreion is more pronounced earlier in the EoR, likely due to the initially small sizes of bubbles early on in reionization. By the time a large fraction of the simulation volume becomes ionized, this time lag becomes much smaller on average. This is because newly ionized regions are more likely to immediately coalesce into large, pre-existing bubbles or otherwise flash ionize, thereby reducing the time required to reach the specified $R_\text{eff}$ values.

To further quantify the bubble growth after the initial reionization of the region, we study the times and rates at which bubbles double in radius. Bubbles which have become ionized at different redshifts during the EoR have unique bubble growth properties. We specifically look at the doubling time (time for the bubble radius to double in size), growth rate (change in effective volume over doubling time), and effective speed of ionization fronts (change in effective radius over doubling time) in Figure~\ref{fig:doubling_time}. Each colour point indicates the radius change we are considering (i.e. the dark purple shows the doubling times and growth rates for bubbles growing from 8 cMpc in radius to 16 cMpc). The horizontal axis is \zreion allowing us to view how different populations' bubbles grow differently, whether they were ionized early in reionization or later. We note that both growth rate and ionizing front speed are calculated assuming spherical bubbles, and thus, may not fully represent the morphological complexities present once bubbles coalesce.

By inspection of Figure~\ref{fig:doubling_time}, we see that the latest ionized regions on the left side of the plots have the shortest doubling times and, consequently, the fastest growth rates. This is consistent with our picture of ``flash-ionization'' of regions late in the EoR as they quickly join large bubbles after becoming ionized, i.e. runaway percolation. The doubling times for regions ionized before $z \approx 12$ are similar and quite high, around 100\,Myr, even for the different bubble radii. This could be partially due to the long time between local reionization in the region and the end of the EoR, but may also indicate exponential behavior in the growth rate of the radius, as the characteristic doubling time appears to be independent of the bubble radius.

\section{Conclusions}
\label{sec:conclusion}

One of the most promising methods to study the large-scale processes during the EoR is the topological analysis of ionized bubbles. A significant body of work has been dedicated to developing procedures for identifying and characterizing these bubbles from idealized calculations, semi-analytic models, and fully-coupled radiative transfer simulations. In this context, this work aims to study ionized bubble size statistics during reionization as modeled by the \thesan simulations. We primarily employ the MFP method, which calculates bubble sizes by extending rays from ionized cells until they intersect with neutral cells, as an effective technique for determining ionized bubble sizes. We examined the time evolution of the bubble sizes, connections between sizes and environmental properties, and the utility of \zreion as an effective proxy for tracking bubble size characteristics.

Our detailed analysis of the \thesan simulations has led to several insights into the nature of ionized bubbles during the EoR. We summarize our main findings as follows:
\begin{enumerate}
    \item Bubbles begin to form coincident with the first significant ionizing sources around redshift $z \sim 15$ and exhibit a growth trajectory from $R_\text{eff} \sim 100\,\text{ckpc}$ to $\sim 100\,\text{cMpc}$ by redshift $z \sim 5.5$, indicative of the progressive completion of cosmic reionization within the simulation volume.
    \item Bubble sizes are notably larger around bright galaxies. This suggests that observational measurements of bubble sizes around high-redshift sources are likely skewed towards larger sizes, and accounting for this bias helps reconcile the discrepancy between predicted and observed bubble sizes.
    \item We find redshift-dependent variations in bubble growth rates. Specifically, regions that undergo reionization early in the EoR display relatively slow bubble expansion. In contrast, regions that are reionized later in the EoR experience ``flash-ionization'' and rapidly join larger, pre-existing bubbles.
    \item Bubbles in high-density regions are generally larger than those in low-density regions, particularly when $x_{\HI} \gtrsim 0.5$ in the earlier stages of the EoR. This correlation is most pronounced at very high overdensities before the bubbles have expanded and merged enough to wash out correlations with environment. A similar trend exists for bubbles surrounding massive galaxies. In both cases, we characterize the median and variance in bubble sizes, and show the power-law slope $d\log R_{\rm eff}/d \log M_{\rm halo}$ flattens as reionization progresses.
    \item We comment on the utility of the redshift of reionization (\zreion) as a viable theoretical tracer for small bubble sizes, particularly for radial sizes below approximately $\lesssim 1\,\text{cMpc}$. This provides a convenient local metric for studying non-local phenomena, and capturing the temporal history of ionized regions. We note that at high redshifts, there exists a substantial time lag between local reionization and large bubble growth, but by $z \lesssim 6.5$, this delay significantly shortens.
\end{enumerate}

In summary, this work enhances our understanding of the topological and evolutionary properties of ionized bubbles during the EoR, including connections between bubble sizes and both small-scale galaxy and large-scale environmental properties.

There is also a wealth of new observational data available from high-$z$ galaxy spectra from the \textit{JWST}. This data can be harnessed to constrain both ionized bubble sizes and the neutral fraction through the analysis of Ly$\alpha$ damping wings. Thus, it would be optimal to identify bubble sizes through a synthetic observation framework that models the spectra and data reduction pipeline currently in use. It would be natural to extend this study to incorporate similar systematics as the observations and facilitate more complete physical connections to theoretical topological metrics.

Further follow-up studies to this could also include performing a more extensive analyses of bubble sizes across the other simulations in the \thesan suite, which include different dark matter models, escape fractions, and dominant sources of ionizing radiation. Such comparative studies would elucidate the distinct signatures imprinted by different physical processes within the ionized bubbles. Moreover, we plan to perform a complementary topological examination of \zreion to analyze the formation and evolution of ionized regions, which will provide invaluable information about the mechanisms by which the Universe became ionized. Follow-up work will aim to continue to draw conclusions about the interplay of reionization and galaxy formation, including both local and environmental information.

\section*{Acknowledgements}
We thank the anonymous referee for their insightful comments and suggestions. We thank Koki Kakiichi and Laura Keating for insightful discussions related to this work.
AS acknowledges support under an Institute for Theory and Computation Fellowships at the Center for Astrophysics $\vert$ Harvard \& Smithsonian.
MV acknowledges support through NASA ATP grants 19-ATP19-0019, 19-ATP19-0020, 19-ATP19-0167, and NSF grants AST-1814053, AST-1814259,  AST-1909831 and AST-2007355.
EG acknowledges support from the CANON Foundation Europe through the Canon Fellowship program during part of the work presented in this paper.
The authors gratefully acknowledge the Max Planck Computing and Data Facility (\url{https://www.mpcdf.mpg.de/}) for support in hosting data and releasing them to the public, as well as the Gauss Centre for Supercomputing e.V. (\url{www.gauss-centre.eu}) for funding this project by providing computing time on the GCS Supercomputer SuperMUC-NG at Leibniz Supercomputing Centre (\url{www.lrz.de}).
Additional computing resources were provided by
the Engaging cluster supported by the Massachusetts Institute of Technology.
We are thankful to the community developing and maintaining software packages extensively used in our work, namely: \texttt{matplotlib} \citep{matplotlib}, \texttt{numpy} \citep{numpy}, and \texttt{scipy} \citep{scipy}.


\section*{Data Availability}
All data produced within the \thesan project are fully and openly available at \url{https://thesan-project.com}, including extensive documentation and usage examples \citep{Garaldi2023}. We invite inquiries and collaboration requests from the community.
In conjunction with this paper we add effective bubble size grid calculations to our online repository. Furthermore, a minimal version of our parallelized C++ ray-tracing MFP code is accessible at \url{https://github.com/meredithneyer/mfp-bubbles-thesan}.




\bibliographystyle{mnras}
\bibliography{biblio}




\appendix

\section{Impact of grid resolution}
\label{appx:res}
\begin{figure}
    \centering
    \includegraphics[width=\columnwidth]{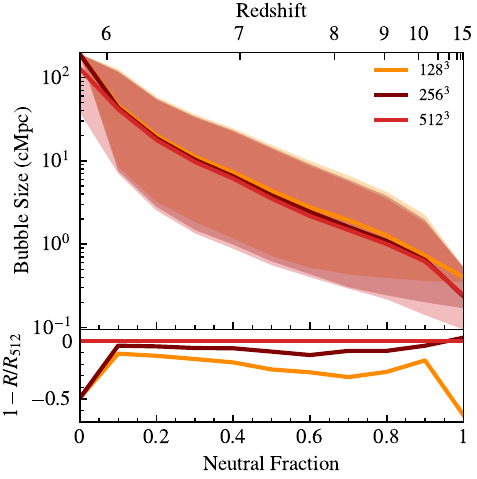}
    \caption{\textit{Upper panel:} Mean-free path determined median bubble size for \thesanone for three resolutions: $128^3,\ 256^3,\ 512^3$. The shaded areas show one standard deviation ranges. \textit{Lower panel:} Relative difference compared to the highest resolution run, $1-R_N/R_{512}$, showing the bubble sizes are converged to within 10 per cent accuracy.}
    \label{fig:sizes_resolution}
\end{figure}
In this appendix, we examine the effects of simulation resolution on ionized bubble size estimates for resolutions of 128, 256, and 512 cells per side. The \thesan simulations and \areport code employ a Voronoi tessellation constructed from a particle representation of the underlying gas distribution. Cartesian grid outputs are subsequently produced by translating these particle locations and properties onto a three dimensional cubic grid with 1024 cells per side. For the purposes of this study, we downsample this grid to produce lower resolution outputs at 512, 256, and 128 cells per side. Due to computational constraints, we utilize the 512 cells per side output as the fiducial highest resolution for this work. We demonstrate convergence of bubble size estimates between the lower resolution runs and the 512 cells per side resolution, thereby confirming that the effects of resolution on our findings are negligible.

The bubble size distributions for each of the resolutions look qualitatively similar. Specifically, the bubble sizes grow from $\sim 100$\,ckpc scales at the onset of reionization to $\sim 100$\,cMpc scales by the end of reionization.

We compare the time evolution of the characteristic bubble sizes across different resolution outputs to check convergence. Figure \ref{fig:sizes_resolution} shows the median bubble sizes as a function of neutral fraction for resolutions of 128, 256, and 512 cells per side. We find that bubble sizes are converged to within approximately 10 and 30 per cent accuracy for 256 and 128 cells per side resolutions, respectively. The bubble sizes of the lower resolution simulations tend to be larger than the bubble sizes for the 512 cells per side run because the cells themselves are larger and artificially expand the bubbles. 

The bubble sizes primarily diverge at the very beginning and end of reionization. This is expected due to the resolution-scale effects becoming important when there are only small patches of ionized and neutral gas at $x_{\HI} \gtrsim 0.9$ and $x_{\HI} \lesssim 0.1$, respectively. At the beginning of reionization, small pockets of ionized gas are either washed out or blurred into an artificially large bubble on the scale of the cell size, which will lead to more divergence from the high resolution model than intermediate times when the scales of neutral and ionized gas regions are much larger than the cell sizes. Similarly, at the end of reionization when the last 10 per cent of gas is becoming ionized, there are small-scale regions of neutral gas remaining in the high resolution 512 cell per side box, but those have been washed out by the resolution effects leading to a larger bubble size by the end of reionization.

\section{Impact of grid smoothing scale}
\label{appx:smooth}
\begin{figure}
    \centering
    \includegraphics[width=\columnwidth]{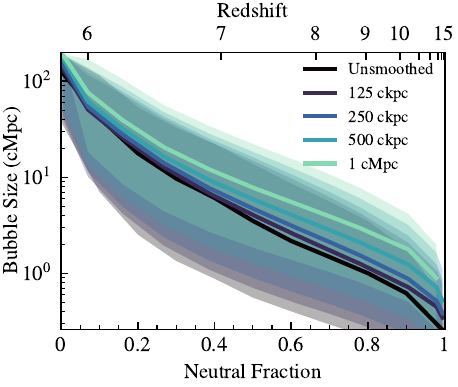}
    \caption{Median bubble sizes over time as characterized by neutral fraction and redshift for different smoothing scales. The smoothing scales do not seem to change bubble sizes much at later times, but bubbles are biased larger for larger smoothing scales, early on. This is partly due to the blurring of small ionized regions during the smoothing process which artificially increases bubble sizes, likely on the order of the smoothing scale. Once many bubbles are larger than the smoothing scale, the bias introduced by the smoothing becomes negligible.}
    \label{fig:sizes_smoothing}
\end{figure}

We also investigate the effects of smoothing on the derived bubble sizes. The smoothing we use in this work, primarily for environmental effects is volume-weighted. The smoothing is performed using a Gaussian kernel applied to the simulation outputs before converting to a Cartesian grid.

We study the behaviour of the characteristic bubble sizes for several different smoothing scales: unsmoothed, 125\,ckpc, 250\,ckpc, 500\,ckpc, and 1\,cMpc. The comparison of their median bubble sizes and one sigma regions are shown in Figure \ref{fig:sizes_smoothing}. As the smoothing scale increases, so does the median bubble size. This is likely due to the blurring of ionized regions with the smoothing leading to a larger observed bubble size. At the beginning of reionization when the universe is still over 90 per cent neutral, the median bubble sizes depend heavily on the smoothing scale. When the first ionized bubbles begin to form, they are immediately blurred by the smoothing kernel leading to the smallest bubble sizes being roughly the size of the smoothing scale (and for the unsmoothed case, the smallest bubbles are approximately the size of one cell). These smoothing-dependent effects appear to wash out as reionization progresses, likely due to most ionized cells being part of bubbles with sizes much larger than the smoothing scales. By the time reionization is ending, smoothing effects on the median bubble size seem to have nearly completely disappeared as the entire simulation box becomes ionized and part of one large bubble, as expected.

We explore the effects of smoothing for two main reasons: the environmental analysis and future computational convenience. In order to quantify the effect of the environment on one simulation cell's bubble size, we need to smooth the density to get information about the surrounding medium in the simulation. Checking that the bubble sizes are consistent under this smoothing ensures that our environmental results reflect the conditions in the local surrounding areas of the bubble centres without artificially impacting the bubble size measurements.

We also check the effects of smoothing on bubble sizes to inform future work that may take advantage of the computational convenience of having a smoother field to analyze. Since many topological analyses require smooth fields, it is important for future work to be able to quantify the effects of this smoothing on bubble sizes.

\section{Impact of simulation resolution}
\label{appx:conv}
\begin{figure}
    \centering
    \includegraphics[width=\columnwidth]{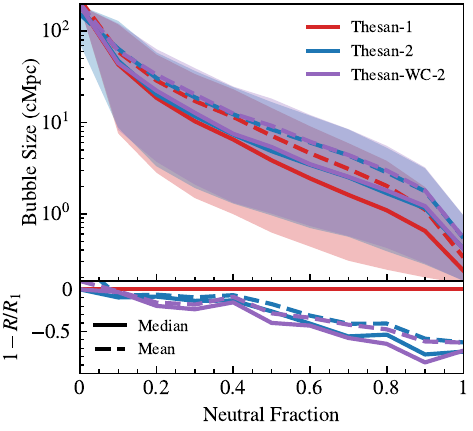}
    \caption{Median (solid) and mean (dashed) bubble sizes over time as characterized by neutral fraction and redshift for different simulation resolutions. The statistics are affected most strongly during the first half of reionization ($x_{\HII} \lesssim 0.5$) due to the relatively numerous small bubbles around the lowest-mass haloes resolved by the simulation and escape fraction differences. The agreement improves as bubbles coalesce into larger ones.}
    \label{fig:sizes_conv}
\end{figure}

In this appendix, we demonstrate the convergence of bubble sizes with simulation resolution. Specifically, in Fig.~\ref{fig:sizes_conv}, we analyze the median and mean $R_\text{eff}$ evolution from the \thesantwo and \thesanwc medium resolution simulations in comparison with the results from the flagship \thesanone simulation. Recall from Section~\ref{sec:environment} that these are both from the same initial conditions as \thesanone but with two (eight) times lower spatial (mass) resolution, and that \thesanwc has a increases the birth cloud escape fraction from 0.37 to 0.43 compensate for lower star formation. We note that these calculations employ a grid resolution of $256^3$, which has been shown in Fig.~\ref{fig:sizes_resolution} to be converged with the $512^3$ results within 10 per cent at all times. The impact on the median and mean statistics is most pronounced in the initial stages of reionization, primarily due to the abundance of small bubbles surrounding the lowest-mass haloes captured by the \thesanone simulation, as well as variations in escape fraction \citep[see][]{Yeh2023}. As these bubbles merge to form larger structures, the statistical agreement improves.


\bsp	
\label{lastpage}
\end{document}